# Concept driven framework for Latent Table Discovery

Gowri Shankar Ramaswamy and F Sagayaraj Francis

**Abstract**—Database systems have to cater to the growing demands of the information age. The growth of the new age information retrieval powerhouses like search engines has thrown a challenge to the data management community to come up with novel mechanisms for feeding information to end users. The burgeoning use of natural language query interfaces compels system designers to present meaningful and customised information. Conventional query languages like SQL do not cater to these requirements due to syntax oriented design. Providing a semantic cover over these systems was the aim of latent table discovery focusing on semantically connecting unrelated tables that were not syntactically related by design and document the discovered knowledge. This paper throws a new direction towards improving the semantic capabilities of database systems by introducing a concept driven framework over the latent table discovery method.

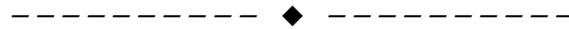

## 1 INTRODUCTION

DATA storage and Information retrieval is ubiquitous nonetheless are dependent on underlying formats based on the application. A global query interface like a search engine is left in a conundrum if it has to manage a multitude of repositories each having different structure and format. Providing semantic interoperability is the way forward. The challenge in semantics is to precisely connect related terms. Since different conceptualisations of the world exist among people due to linguistic, cultural, religious, geographic, demographic differences, there is a need to document conceptual level knowledge tangibly so that information systems can produce customised results. Making systems intelligent involves a plethora of challenges most importantly modeling intelligence to replicate human style of thinking. Cognitive science [1] has spearheaded the goal of understanding how humans think and has significantly enriched the literature with findings on knowledge, knowledge acquisition and representation by human beings, human cognition, human way of reasoning logic, application of captured knowledge to problem solving, memory retention and retrieval and so forth. Cognitive psychology [2] has also contributed copiously by studying mental processes and the behavior of the human mind. These findings are guidelines to computer scientists working towards incubating intelligence into the computer. Yet there are prodigious roadblocks to adapting these findings into the domain of computer science. For example, human beings are subjective whereas computers are objective. Everything within a computer works on predefined rules, logic and algorithms but human beings tend to override rules, logic based on needs and circumstances. Since computing systems trace their roots to mathematical models, everything about human beings that has been discovered by the cognitive fields must be mapped to mathematics which is a mammoth effort per se. This paper takes one such issue that is discovering meaningful relationships between data entities inside data sources and adapts a concept [3] based model to improve the relationship discovery process. In this paper, the term database means a collection of data in a logical and structured format. Concept is a manifestation of elements in the real world with characteristics and behavior. Knowledge is the familiarity of concepts, facts, information, descriptions, truths, principles acquired through experience and education. Ontology is a representation of set of concepts within a domain and relationships between them. Semantics is the meaning of things and the study of meaning. Semantic interoperability in is the ability to exchange data with unambiguous and shared meaning.

## 2 LATENT TABLE DISCOVERY (LTD)

LTD states that:

> *Given a set of entity sets, LTD discovers other closely related entity sets that are nonexistent in the set of entity sets earlier by taking into account their co-occurrence in domain ontology* [4]

Tables within relational databases are linked to one another only through referential integrity constraints. When a query requires data from multiple tables, joins have to be performed to combine the data which is a costly operation per se. The database designer provides refe-

- *Gowri Shankar Ramaswamy is a PhD Scholar in Computer Science & Engineering at Pondicherry Engineering College, Puducherry, India.*
- *F Sagayaraj Francis is an associate professor in the department of Computer Science & Engineering at Pondicherry Engineering College, Puducherry, India.*

rential constraints at the time of design only on tables that need to be joined by envisaging the possible queries on the system. The gamut of queries keeps increasing with time so a syntactic method to connect tables is not scalable. LTD positions itself in such scenarios by exploring semantic connections between tables that are hitherto unconnected.

In mathematical terms, we have unrelated relational tables with K set of tuples and Y set of attributes. Input is two such tables $T_1$ and $T_2$ and a correspondence condition between any one attribute $Y_i$ of $T_1$ and $Y_j$ of $T_2$, an ontology containing the domain knowledge of all the attributes of the tables represented as $O = \{C, L\}$ where C is a set of concepts and L is a set of links between the concepts is also considered. The method to find out the link between these tables is:

1. The ontology is traversed to find semantic match with $Y_1$ as well as with $Y_2$ through one or more intermediate concepts. By semantic match we mean a match derived out of thesauruses that point to the concept in contention.
2. A new relation is excavated as shown in (1)

$$Y_i \xrightarrow{X} Y_j \qquad (1)$$

where $Y_i, Y_j \in Y$ and $i, j = 1$ i.e. LTD finds a link between one attribute from $T_1$ and one attribute from $T_2$;
X is a concept connecting $Y_i$ with $Y_j$ semantically.

Observe that we have made a transitive link between $Y_i$ and $Y_j$ via X. Also, if X is set of concepts then this becomes a transitive closure of the Ontology formally written in (2)

$$Y_i \xrightarrow{X^*} Y_j \qquad (2)$$

This discovery is thus a semantic transitive closure.

Given a set of tables, ontology of the system and a correspondence between the columns of the tables to be checked for semantic links, LTD with the help of the ontology finds connections between the columns and documents the newly discovered relations again as tables.

The proof of feasibility and the complexity of LTD are discussed in [4].

## 3 LTD WITHOUT CONCEPTUAL KNOWLEDGE – A CASE STUDY

Consider two unrelated tables within a database namely Diseases table shown in fig. 1 and States table shown in fig. 2 within a database.

DISEASES TABLE

| Disease Name | Disease Fatality |
|---|---|
| Fever | Low |
| Paralysis | High |
| Jaundice | Medium |
| Chicken Pox | High |

Fig. 1. Data contained in a table about diseases

STATES TABLE

| State | Capital City |
|---|---|
| Rajasthan | Jaipur |
| Karnataka | Bangalore |
| Himachal Pradesh | Shimla |
| Puducherry | Puducherry |

Fig. 2. Data contained in a table about states

Consider the domain ontology of seasonal diseases shown in fig. 3

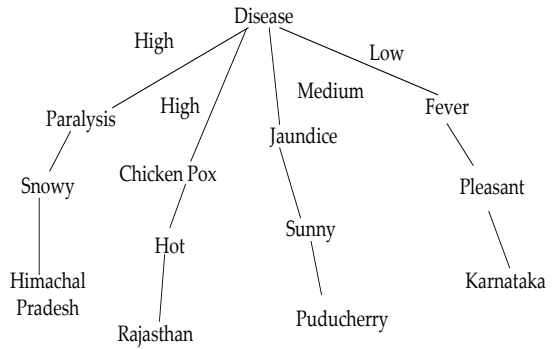

Fig. 3. Domain Ontology of Seasonal Diseases for the database containing diseases and state table

If the input correspondence condition to LTD is between Disease Name and State Name, the outcome of LTD is as shown in fig. 4.

RESULTANT TABLE OF LTD WITHOUT CONCEPTUAL KNOWLEDGE

| Disease Name | Weather | State Name |
|---|---|---|
| Jaundice | Sunny | Puducherry |
| Fever | Pleasant | Karnataka |
| Paralysis | Snowy | Himachal Pradesh |
| Chicken Pox | Hot | Rajasthan |

Fig. 4. Data contained in the resulting table after the LTD process without using conceptual knowledge

Disease name is connected to State name via weather. This table can be interpreted to give information on the states that are likely to be infected by diseases based on the prevailing weather forecast.

## 4 CONCEPTUAL SPACE

Peter Gärdenfors in his paper published in Mind and Matter [3], claims that, prior to his approach on conceptual space, in cognitive science, there are two approaches to model representations namely 1) symbolic approach which says that cognition is a manipulation of symbols, and 2) associative approach which says that the representation of the association between information elements. Peter also claims that both these models, one that deals with symbol and other that represents connection in an artificial neural network are not sufficient for representing knowledge as they approach the problem from a much lower level than how human beings visualise knowledge and proposed a third approach namely conceptual space model. The conceptual way involves geometric representation of concepts based on a number of quality dimensions [3]. One of primal tasks in cognitive processes is the judgment of similarity. Symbolic representations largely inhibit the notion of computing similarity between elements. Objects were mapped into Euclidean space and the distance between objects in that space was used to ascertain similarity. The smaller the distance between objects, greater is their similarity. The introduction of conceptual space is a significant step towards effectively evaluating similarity from an epistemological perspective since it is able to add a new dimension in the form of meaning. Another important role of conceptual spaces is to accommodate different perceptions of relation between objects that can vary with time or context. Quality dimensions [3] capture the qualities of an object. For example, temperature, height, weight, width, depth. Quality dimensions can also be abstract and derived. A point or mapping on the conceptual space would have points representing an instance of each quality dimension; say for example, a particular height or weight. Raubal [5] based his work on formalizing conceptual spaces in the field of geography in the approach propounded by Gärdenfors [3]. Raubal [5] argued that concepts are mental entities that capture experience and support reasoning of the world. He also claimed that concepts are not static but are dynamic as they are capable of changing with time. In this paper, an adaption of conceptual space is presented in the field of data management to add value to the services offered by data repositories. A conceptual cover over the afore-mentioned latent table discovery method is shown to enhance the quality of results produced.

## 5 CONCEPTUAL FRAMEWORK IN LTD

Latent table discovery aimed at extracting new relationships between database tables that were hitherto not explicitly marked as related by the database design. LTD, though was shown to be feasible, was also observed to be with exponential time complexity and was impossible to solve in real time using a computer [3]. Nevertheless, ideal cases were also identified where LTD could provide computable results. LTD requires an ontology to make decisions and the computability of LTD is influenced by its ability to identify meaningful links between the entity sets of the tables currently under examination. A conceptual intelligence about the domain of the entity sets is therefore necessary. By providing a conceptual framework, each term in the domain of the entity sets is viewed as a concept with certain characteristics. Those terms in the domain with one or more similar characteristics can be classified together so that LTD can establish a connection between them and excavate new relations.

In mathematical terms, we have unrelated relational tables with K set of tuples and Y set of attributes. Input is 1) Set of unrelated tables T, 2) ontology containing the domain knowledge of all the attributes of the tables represented as $O = \{C, L\}$ where $C$ is a set of concepts and $L$ is a set of links between the terms, 3) Conceptual Knowledge $C_n$ = <$D_1, D_2, …, D_n$> where $C_n$ is the set of concepts of the domain and $D_n$ are the quality attributes or quality dimensions of the concepts. If the quality dimensions represent a domain, then $D_n$ = <$d_1, d_2, …, d_n$>. The method to find out the link between these tables is:

1. The Ontology is traversed to find semantic match between the Y set of attributes of the T set of tables through one or more intermediate concepts. The conceptual knowledge is used to obtain the definition of the concepts and those with matching quality dimensions are linked.
2. Newly found relations are of the form written in (3)

$$Y_i \xrightarrow{X^*} Y_j \qquad (3)$$

where $Y_i, Y_j \in Y$ and $i, j \geq 1$ i.e. LTD with conceptual knowledge can link any number of attributes from any number of unrelated tables;
$X^*$ is the set of intermediate concepts connecting $Y_i$ and $Y_j$.

This discovery is also a semantic transitive closure.

## 6 LTD WITH CONCEPTUAL KNOWLEDGE – A CASE STUDY

Consider a partial conceptual knowledge of the system as described below:

A concept in conceptual space is any worldly entity that it is described based on properties. For example, concept 'weather' is defined based on the property 'temperature'. Concepts can also be defined by nesting other defined concepts. For example, concept 'weather' is a collection of nested concepts namely 'snowy', 'cold', 'pleasant', 'sunny' and 'hot'; each of them having been defined based on the property 'temperature'. Concept 'place' has properties 'name', 'state', 'weather', 'zone' and 'population'. Such a conceptual knowledge gives clarity to the meaning of the individual entities of the tables and the elements in the ontology based on which LTD can excavate new relations.

Weather = <snowy, cold, pleasant, sunny, hot>

Snowy: {-10 °C < temperature < 10 °C}
Cold: {11 °C < temperature < 20 °C}
Pleasant: {11 °C < temperature < 22 °C}
Sunny: {23 °C < temperature < 35 °C}
Hot: {36 °C < temperature < 50 °C}

Place = <name, state, climate, zone, population>
Sample domain of Place: {
(Mandi, Himachal Pradesh, Snowy, North, 155023),
(Ooty, Tamil Nadu, Cold, South, 345426),
(Mysore, Karnataka, Pleasant, South, 1242560),
(Puducherry, Puducherry, Sunny, East, 1175236),
(Thar, Rajasthan, Hot, West, 26756)
}

Suppose that, apart from Diseases table shown in fig. 1 and States table shown in fig. 2, the database has two more tables namely Daily Temperature of places shown in fig. 5 and List of Zonal Health Monitoring Centres shown in fig. 6. With the introduction of the conceptual knowledge, LTD can fine-tune its discovery by studying the domain of the weather and places based on their definion in the knowledge and can find new relations like the one given in fig. 7.

DAILY TEMPERATURE TABLE (RECORDED ON SOME PARTICULAR DAY)

| Place | Temperature (in °C) |
|---|---|
| Mandi | 10 |
| Thar | 42 |
| Puducherry | 33 |
| Mysore | 20 |

Fig. 5. Data contained in a table about daily temperature recording of places

ZONAL HEALTH MONITORING CENTRES TABLE

| Zone | Centre |
|---|---|
| North | New Delhi |
| East | Chennai |
| West | Mumbai |
| South | Bangalore |

Fig. 6. Data contained in a table about the list zonal health monitoring centres

The outcome of LTD with conceptual knowledge is shown in fig. 7.

RESULTANT TABLE OF LTD WITH CONCEPTUAL KNOWLEDGE

| Disease Name | Weather | Place | Population | Health Monitoring Centre |
|---|---|---|---|---|
| Jaundice | Sunny | Puducherry | 1175236 | Chennai |
| Fever | Pleasant | Mysore | 1242560 | Bangalore |
| Paralysis | Snowy | Mandi | 155023 | New Delhi |
| Chicken Pox | Hot | Thar | 26756 | Mumbai |

Fig. 7. Data contained in the resulting table after the LTD process using conceptual knowledge

With the help of the conceptual knowledge, LTD could excavate a far more refined relation that projects the diseases that are likely to affect the places, the population expected to be affected and also provides information about the nearest Health Monitoring Centre that has to be alerted in case of an outbreak. As it can be noted, the introduction of conceptual knowledge has obviated the need to explicitly mention the columns of tables between which the existence of a semantic connection is to be determined by LTD. Also, this new relation can be used to enrich the ontology of the system with more details.

## 7 CONCLUSION AND FUTURE WORK

Semantics is the intention behind practicing conceptual modeling or use of the concept model after the conceptualization process. Hence to enable data sources to support semantics based processing, conceptualization has to be first introduced which is suggested in this work. The conceptual model followed in this paper uses a vector algebraic method which was proposed by Raubal et al [5]. The representation of the concepts, per se, poses lot of challenges including:

1. Prior research in this front has proposed representing conceptual spaces using the theory of vector space [6]. Here a concept is defined in terms of a set of quality attributes or dimensions. For example, a concept named 'human' may have quality attributes: height, weight, gender. These quality attributes are assumed to be mutually exclusive which may not be the case always [7]. Example: consider attributes weight, waist size of the concept 'human'. A reduction in weight can also alter waist size and vice versa. In such cases, there a need to analyse covariance between dimensions [8].
2. Knowledge represented at the conceptual level must be capable of being mapped into one of existing semantic technologies (like OWL ontologies) to be of use in the semantic web. This poses a lot of challenges at the mapping level per se [9].
3. At a fundamental level, if vector spaces are used to represent knowledge [3], the tacit implication is conceptual elements are linear. When the real world is a non-linear system, it is imperative to investigate the likelihood of conceptual spaces being non-linear.
4. If conceptual spaces are presumed to be non-linear, the

possibility of being chaotic must also be explored
5. Concepts change and evolve with time. Cardinality of the attributes of a concept also changes with time [7]. This dynamism must be appropriately addressed.

Future work intends to solve these challenges.

**Gowri Shankar Ramaswamy** is a PhD Scholar in the Department of Computer Science & Engineering at Pondicherry Engineering College, Puducherry, India. He received his M.Tech (2011) and B.Tech (2007) from Pondicherry University, Puducherry, India. He has published 1 research paper in an International Conference and 1 paper in a Journal. His research interests include Knowledge Discovery, Data Semantics, Data Management.

**F Sagayaraj Francis** holds a PhD in Data Management from Pondicherry Univeristy, Puducherry, India. He is an associate professor in the Department of Computer Science & Engineering at Pondicherry Engineering College, Puducherry, India. He has to his credit 6 journal and 7 conference publications. His research interests include Data Management, Geographic Information Systems.
.